\documentclass[useAMS,usenatbib]{mn2e} 
\usepackage{graphicx}
\usepackage{url}

\newcommand{\MSun}{\mbox{${\rm M}_\odot$}}
\newcommand{\Msun}{\mbox{${\rm M}_\odot$}}

\def\lteq{\ {\raise-.5ex\hbox{$\buildrel<\over-$}}\ }
\def\apgt{\ {\raise-.5ex\hbox{$\buildrel>\over\sim$}}\ }
\def\aplt{\ {\raise-.5ex\hbox{$\buildrel<\over\sim$}}\ }
\def\lt{\ {\raise-.5ex\hbox{$\buildrel>$}}\ }
\def\gt{\ {\raise-.5ex\hbox{$\buildrel<$}}\ }
\def\eqgt{\ {\raise-.5ex\hbox{$\buildrel>\over-$}}\ }
\def\eqlt{\ {\raise-.5ex\hbox{$\buildrel<\over-$}}\ }

\newfont{\Giga}{cmssbx10 scaled 5200}
\newfont{\giga}{cmssbx10 scaled 4300}
\newfont{\Mega}{cmssbx10 scaled 3200}
\newfont{\mega}{cmssbx10 scaled 2500}
\newfont{\Kilo}{cmssbx10 scaled 2000}
\newfont{\kilo}{cmssbx10 scaled 1600}
\newfont{\Deca}{cmssbx10 scaled 1450}
\newfont{\deca}{cmssbx10 scaled 1200}
\newfont{\Dezi}{cmssbx10 scaled 1100}
\newfont{\dezi}{cmssbx10 scaled 1050}
\newfont{\iGiga}{cmssi10 scaled 6200}
\newfont{\igiga}{cmssi10 scaled 4300}
\newfont{\iMega}{cmssi10 scaled 3200}
\newfont{\imega}{cmssi10 scaled 2500}
\newfont{\iKilo}{cmssi10 scaled 2000}
\newfont{\ikilo}{cmssi10 scaled 1500}
\newfont{\mathGiga}{cmsy10 scaled 6200}
\newfont{\mathgiga}{cmsy10 scaled 4300}
\newfont{\mathMega}{cmsy10 scaled 3200}
\newfont{\mathmega}{cmsy10 scaled 2500}
\newfont{\mathKilo}{cmsy10 scaled 2000}
\newfont{\mathkilo}{cmsy10 scaled 1500}
\newfont{\mathDeca}{cmsy10 scaled 1450}
\newfont{\mathdeca}{cmsy10 scaled 1200}

\def\aap{\ {A\&A}\ }

\def\apj{\ {ApJ}\ }
\def\apjl{\ {ApJL}\ }

\def\araa{\ {ARA\&A}\ }

\def\icarus{\ {Icarus}\ }

\def\mnras{\ {MNRAS}\ }
\def\nat{\ {Nat}\ }

\def\apgt{\ {\raise-.5ex\hbox{$\buildrel>\over\sim$}}\ }
\def\aplt{\ {\raise-.5ex\hbox{$\buildrel<\over\sim$}}\ }
\def\lteq{\ {\raise-.5ex\hbox{$\buildrel<\over-$}}\ }

\title[]{The fragility of planetary systems}

\author[]{S.F. Portegies Zwart$^{1}$ and Lucie J\'{i}lkov\'{a}\\
Leiden Observatory, Leiden University, PO Box 9513, 2300 RA, Leiden, 
The Netherlands }
\begin{document}

\date{}
\maketitle

\begin{abstract} 
We specify the range to which perturbations penetrate a planetesimal
system. Such perturbations can originate from massive planets or from
encounters with other stars. The latter can have an origin in the star
cluster in which the planetary system was born, or from random
encounters once the planetary system has escaped its parental cluster.
The probability of a random encounter, either in a star cluster or in
the Galactic field depends on the local stellar density, the velocity
dispersion and the time spend in that environment.  By adopting order
of magnitude estimates we argue that the majority of planetary systems
born in open clusters will have a {\em Parking zone}, in which
planetesimals are affected by encounters in their parental star
cluster but remain unperturbed after the star has left the
cluster. Objects found in this range of semi-major axis and
eccentricity preserve the memory of the encounter that last affected
their orbits, and they can therefore be used to reconstruct this
encounter. Planetary systems born in a denser environment, such as in
a globular cluster are unlikely to have a Parking zone. We further
argue that some planetary systems may have a {\em Frozen zone}, in
which orbits are not affected either by the more inner massive planets
or by external influences. Objects discovered in this zone
will have preserved information about their formation in
their orbital parameters.
\end{abstract}


\section{Introduction}

Planetary systems seem to be composed of one or more stars, orbited by
about a dozen planets and many minor bodies
\citep{1632Dialogo...G}. The latter can be roughly divided into
hundreds of moons and dwarf planets, and many millions of
planetesimals.  The objects closer to the star seem to be organized in
a disk-like structure in which also the planets reside, and which
flares to become spherical at larger distance from the stellar host.

This view of planetary systems is heavily based on the Solar System
\citep[see the review][]{2010ARA&A..48...47A}, but so far its
generality cannot be excluded, because each of the several thousands
planetary systems known today \citep{2013Sci...340..572H} seem to
comply to this characteristic.  This view is also supported by our
limited understanding of the formation of planetary systems
\citep{2002ApJ...581..666K,2008ApJ...683..479B,2013ApJ...775...42I}.

The orbits of planets and minor bodies within a few stellar
radii are affected by tidal evolution \citep{1977A&A....57..383Z}. Once
the star leaves the main-sequence, copious mass loss starts to affect
the entire planetary system \citep{2011MNRAS.417.2104V}.  We refrain
from discussing these complexities here, but concentrate on the
dynamically affected regime: sufficiently far away and sufficiently
early in its evolution to remain unaffected by the stellar host.

The effect of perturbations on the Solar System either from the local
planets, external perturbations from the birth cluster or even from
the Galaxy have been studied quite extensively \citep[for a few recent
  studies see e.g.][and references
  therein]{2008Icar..197..221K,2012Icar..217....1B,2014Natur.507..435S}.
We start with a discussion on the Solar System in \S\,\ref{Sect:SS}
and generalize in \S\,\ref{Sect:General}.

\section{Perturbing the Solar System}\label{Sect:SS}

\subsection{The effect of internal perturbations from the planets}

Apart from mass loss and tidal evolution the inner regions of the
Solar System are most strongly affected by dynamical interactions with
the giant planets. We can calculate the range of dynamical
reorganization caused by the widest massive planet in a planetary
system by adopting its apocenter distance. For the Solar System, this
range is currently determined by the planet Neptune, which affects the
orbits of minor bodies to a distance $a \apgt 30$\,AU. Within this
distance the planets in the Solar System have caused major changes in
the orbital distributions of the minor bodies
\citep{2008Icar..196..258L}. We could argue that within a distance of
$30\,\mathrm{AU}/(1-e)$ the minor bodies in the Solar System quickly
lose memory of the mechanism that brought them in these orbits. Here
$e$ is the eccentricity of the planetesimal.


\begin{figure}
\begin{center}
\includegraphics[width=0.5\textwidth]{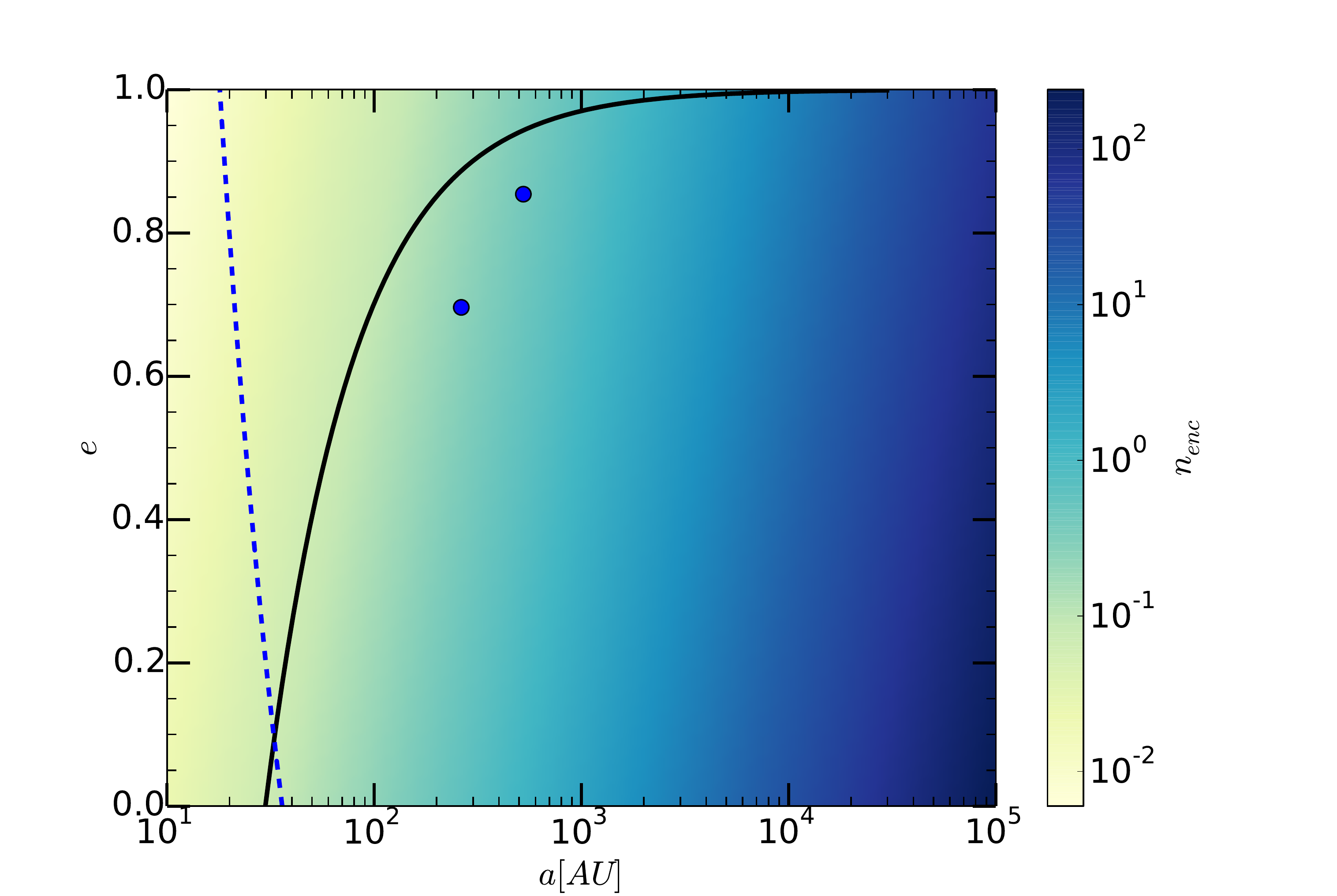}
\end{center}
\caption{Fragility of the Solar System while still a member of its
  birth cluster. The shades represent the probability distribution
  $n_{\rm enc}$, for the number of encounters that perturb the Solar
  System at a given semi-major axis $a$ and eccentricity $e$.  For the
  birth cluster we adopted the mass of $2 \times 10^3$\,\MSun\, and
  virial radius of 2\,pc to remain constant over a time scale of
  200\,Myr. The solid black curve gives the semi-major axis up to
  which Neptune can perturb orbits. The blue dashed curve gives the
  distance to which star Q \citep{Jilkova2015}, that passed the Sun
  with an impact parameter of 320\,AU, has perturbed the Solar System.
  According to the color scaling along the right edge of the figure,
  such a close encounter could have occurred roughly once while the
  Solar System was a member of the star cluster.  The two bullet
  points give the orbital parameters of Sedna
  \citep{2004ApJ...617..645B} and 2012VP$_{113}$
  \citep{2014Natur.507..471T}.
\label{Fig:cluster}
}
\end{figure}

\subsection{The effect of external perturbations from the parental star cluster}

We adopt the view that all stars are born in a clustered environment
\citep{2003ARA&A..41...57L}.  The densities of these environments vary
enormously from $\sim 1$\,star/pc$^3$ to more than $10^6$\,
stars/pc$^3$ \citep{2010ARA&A..48..431P}. The time that the Solar
System with minor bodies remains in the environment determines the
degree by which it is dynamically affected by stellar encounters
\citep[see the review by][]{2006MNRAS.370.2038D}.  Those encounters
play a major role in the evolution of these environments. The degree
by which the planetary system is affected by such encounters depends
on the duration and the intensity of the exposure. Further internal
reorganization of the perturbed planetary system enables the external
perturbations to propagate, generally on a much longer time scale, to
the inner parts of the planetary system.

The way in which a planetary system is affected by dynamical
encounters depends on the mass of the encountering star, its impact
parameter with respect to the host star, the velocity $v$ and the
direction with respect of the planetesimal disk
\citep{2014A&A...565A..32S}.  This complicated combination of
parameters and their mutual relations in terms of the degree of
perturbations for the planetary system can be summarized in a cross
section $\langle \sigma \rangle$.

Most important for preserving the integrity of the planetary system is
its eccentricity.  Here we adopt a lower limit to the perturbation of
the eccentricity of $\delta e = 0.1e$ to be fatal for the particular
planetesimal.  The cross section of eccentricity perturbation of a
planetary system around a star with mass $m$, through an encounter
with another star of mass $M$ has been estimated by means of
integrating small-N systems and averaging over the various encounter
angles \citep{2015MNRAS.448..344L}:
\begin{equation}
  \langle \sigma \rangle = (1-f_b) \langle \sigma_{\rm single} \rangle 
                         + f_b \langle \sigma_{\rm binary} \rangle.
\end{equation}
Here $\sigma_{\rm single}$ is the cross section for encountering a
single star and $\sigma_{\rm binary}$ is for binaries. The parameter
$f_b$ is the binary fraction, which is between 0 and 1. We adopt $f_b
= 0.5$. In the adiabatic regime where the encounter velocity is
comparable to the orbital velocity and which is suitable for encounter
in star clusters, both cross sections have a similar form
\citep{2015MNRAS.448..344L} (with subscript X=single or X=binary
depending on the configuration of the encountering object):
\begin{equation}
  \langle \sigma_{\rm X} \rangle = \sigma_0 {a \over [{\rm AU}]} 
           \left( m \over [\MSun] \right)^{-1/3}
           \left( v \over [{\rm km/s}] \right)^{-\gamma}
           \exp \left( b (1-e_{\rm f}) \right).
\label{Eq:crosssection}
\end{equation}
Here $v$ is the relative velocity for which we adopted the velocity
dispersion in the cluster and $a$ is the orbital semi-major axis of
the planetary system before the encounter.  The post encounter
eccentricity, $e_{\rm f} = e + \delta e$.  Eq.\,\ref{Eq:crosssection}
was calibrated for initially circular orbits, but we apply them here
also for eccentric orbits. According to \cite{2015MNRAS.448..344L} the
cross sections do not depend much on the pre-encounter eccentricity
\citep[but see][]{1996MNRAS.282.1064H}.

The parameters $\sigma_0$, $b$ and $\gamma$ depend on the binarity of
the encountering object. For a single star $\sigma_0 \simeq
1000$\,AU$^2$, $b = 8/5$ and $\gamma = 6/5$, whereas for a binary
$\sigma_0 \simeq 4050$\,AU$^2$, $b = 4/3$ and $\gamma = 7/5$
\citep[see][]{2015MNRAS.448..344L}.  With this cross section the local
stellar density $n$ and the velocity dispersion $\langle v \rangle$ we
can calculate the encounter rate:
\begin{equation}
  \Gamma = n \langle \sigma \rangle  \langle v \rangle.
\label{Eq:Gamma}
\end{equation}

In Figure\,\ref{Fig:cluster} we present the expected value for the
number of encounters in the Sun's parental star cluster.  Here we
adopted the cluster parameters derived by \cite{2009ApJ...696L..13P}:
a total mass of about $M_{\rm cl} = 2\times 10^{3}$\,\MSun,
and a virial radius of 2\,pc, which result in a stellar density of about
100\,stars/pc$^3$ and a velocity dispersion of $\langle v \rangle
\simeq 2$\,km/s.  The lifetime of a star cluster in the Galactic disc
can be estimated from \citep{2005A&A...429..173L}:
\begin{equation}
    t_{\rm cl} = 2.24\, {\rm Myr} \left( {M_{\rm cl} \over [\MSun]} \right)^{0.60}.
\end{equation}
For the star cluster in which the Sun was born this results in a
lifetime of about 200\,Myr.

Our adopted cluster lifetime and the assumption of a constant density
within this period are very approximate. However, our intention
is to estimate the relative importance of experiencing an encounter in
the parental star cluster or in the Milky Way Galaxy, for which this
approach suffices. In order to estimate the importance of these
assumptions we integrated the cluster mass and radius evolution from
the simulations by \cite{2001MNRAS.321..199P}. They aimed their
simulations at mimicking Pleiades, Praesepe and Hyades, which have
comparable initial conditions as the Solar birth cluster. According to
these simulations, clusters with such parameters survive for as long
as a Gyr during which the cluster mass drops linearly with time.  In
this period the cluster expands by about a factor of 3, roughly
proportional to the square root of time. With this slightly more
elaborate estimate we argue that we overestimate the exposure of
encounters in the star cluster by about a factor of two. This would be
consistent with adopting 100\,Myr for the cluster exposure calculation
in Figure\,\ref{Fig:cluster}, rather than 200\,Myr.

The distance to which these perturbations induced by close encounters
penetrate into the Solar System, depends on the distance of closest
approach $q_{\rm enc}$, and the eccentricity of the encounter $e_{\rm
  enc}$, and can be expressed in \citep{2001Icar..153..416K}:
\begin{equation}
    a (1+e) \simeq ({1 \over 5})^{2/3}  q_{\rm enc}
                      \left( (1+M/\Msun) (1+e_{\rm enc}) 
                      \right)^{-1/3}.
\label{Eq:penetration}
\end{equation}

Recently \cite{Jilkova2015} argue that the planetesimals Sedna
\citep{2004ApJ...617..645B} and 2012VP$_{113}$
\citep{2014Natur.507..471T} were captured by a close encounter from a
$M \simeq 1.8$\,\MSun\, star that passed the Solar System with a
closest approach of $q_{\rm enc} \simeq 227$\,AU and relative velocity
$v \simeq 4.3$\,km/s (which corresponds to an eccentricity $e_{\rm
  enc} = 2.6$).  According to Eq.\,\ref{Eq:penetration}, such an
encounter would have perturbed the Solar System to a distance of about
36\,AU.  In Figure\,\ref{Fig:cluster} we present the distance to which
such and encountering star perturbs the planetesimal around the
Sun. For completeness we include the two objects Sedna and
2012VP$_{113}$ to indicate how dramatically the encounter with star Q
perturbed the Solar System.

\subsection{The effect of external perturbations from the Galactic encounters}

Once the parental star cluster dissolves, the planetary system can
only be perturbed by internal reorganization, and by the Galaxy. This
latter can be subdivided in a global perturbation from the slowly
varying Galactic tidal field, but the occasional close encounters with
field stars are more important \citep[see
  however][]{1997Icar..129..106F}.

The probability of an encounter with a Galactic field star is much
smaller than of a close encounter in a star cluster, but the lower
encounter rate is compensated in part by the longer time spent in the
relatively low-density environment of the Galaxy compared to the time
spent in the star cluster.

For estimating the effect of an encounter with a field star we cannot
simply adopt Eq\,\ref{Eq:crosssection} because these are tuned for
low-velocity (and low eccentricity) encounters, whereas Galactic
encounters tend to occur with a much higher velocities.  We therefore
adopt the classic gravitatioal focussed cross section
\citep{1987gady.book.....B}
\begin{equation}
  \sigma = \pi a^2 \left( 1 + {2G(m+M) \over av_{\rm enc}^2} \right),
\label{Eq:GravFocusCrosssection}
\end{equation}
for calculating the encounter rate between the Solar System
($m=1\MSun$) and another Galactic star of mass $M$ to a closest
approach distance $a$.


To provide an upper limit to the effect an encounter has on the
orbital parameters of the planets or planetesimals we assume that the
perturbed object and the closest approach are aligned. The
heliocentric impulse gained by the object at distance $r$ from the Sun
is then given by \citep{1976BAICz..27...92R}:
\begin{equation}
  \Delta v = \frac{2GM}{v_{\rm enc}} \frac{r}{q_{\rm enc}(q_{\rm enc}-r)}.
\label{Eq:impulse}
\end{equation}
Here we assumed that the relative velocity, $v_{\rm enc} = 30$\,km/s,
remains constant during the encounter and the mass of an encountering
star $M = 0.5$\,\MSun.  The perturbation is effective at apohelion
(i.e., $r=a(1+e)$) and we estimate the distance to which the
perturbation penetrates the Solar System at the point where the
impulse gained by the object is comparable to its velocity at
apohelion. This is a rather arbitrary choice, but suffices to indicate
to which distance a passing field star may have affected planetesimals
in the Solar System.

\begin{figure}
\begin{center}
\includegraphics[width=0.5\textwidth]{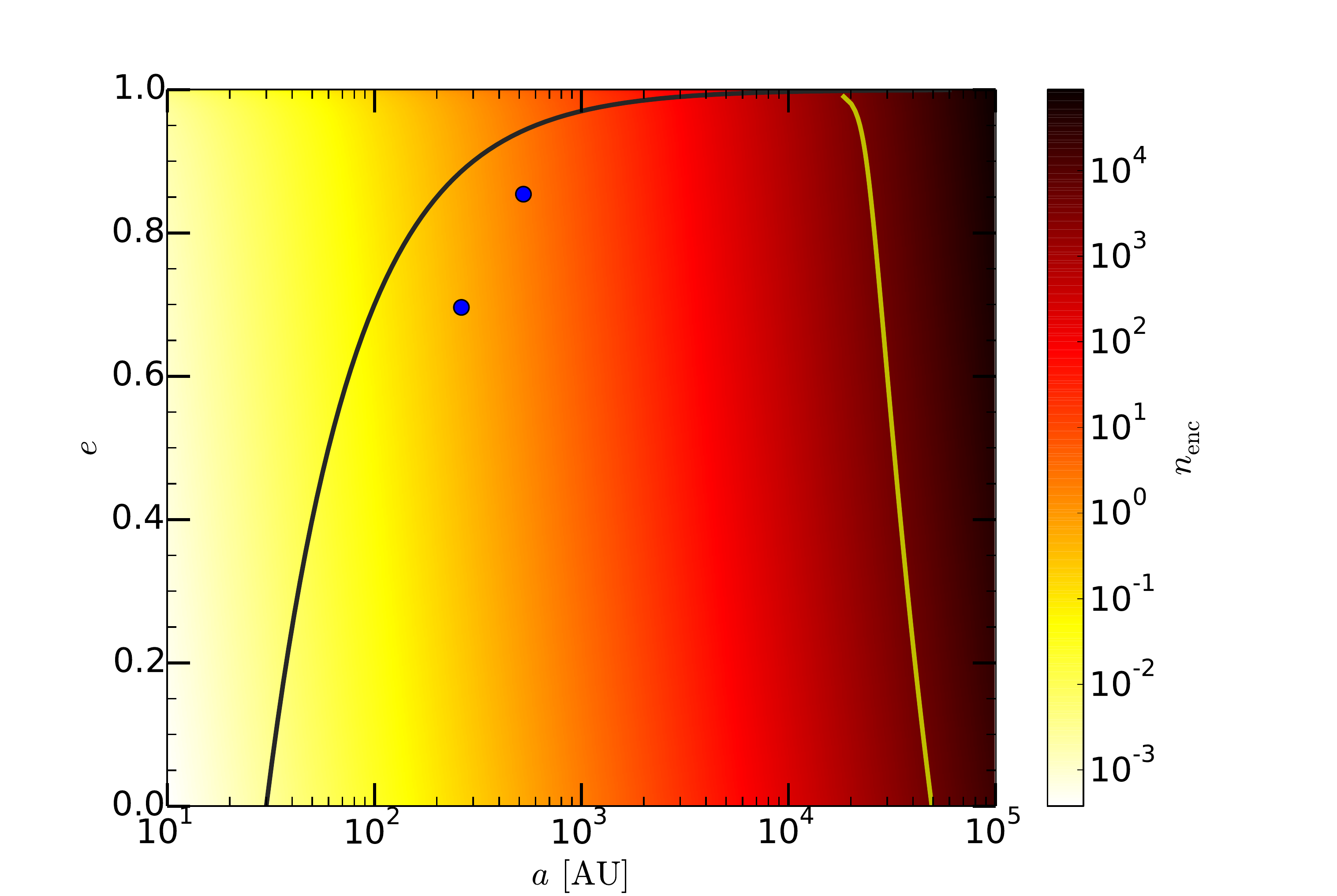}
\end{center}
\caption{The fragility of the Solar System in the Galactic field.  The
  shades represent $n_{\rm enc}$: the probability density distribution
  for the number of encounters that affect the orbits in the Solar
  System to a given semi-major axis $a$ and eccentricity $e$.  Here we
  adopted a stellar density of $n=0.20$\,stars/pc$^3$, a velocity
  dispersion of $v = 30$\,km/s and the typical mass of an encountering
  star of $M = 0.5$\,\MSun.  The solid black curve and the two bullet
  points give the Neptune's perturbing distance, and the orbital
  parameters for Sedna and 2012VP$_{113}$, see also
  Fig.\,\ref{Fig:cluster}.  The yellow curve gives the distance to
  which the Solar System was perturbed (by 1\,\% of its velocity at
  apohelion) due to the recent encounter with the 0.15\,\MSun\, binary
  star WISE~J072003.20-084651.2, which grazed the Solar System, but
  clearly did not come in close enough to perturb the inner Oort
  cloud.
\label{Fig:galaxy}
}
\end{figure}

In Fig.\,\ref{Fig:galaxy} we present the number of encounters that
perturbed the Solar System by a passing Galactic disk star, after the
parental star cluster has been dissolved.  The probability
distribution is calculated using Eq.\,\ref{Eq:Gamma} and adopting $n =
0.20$\,stars/pc$^{3}$ and a velocity dispersion in the local standard
of rest of $v=30$\,km/s \citep{2007A&A...475..519H}, and the time the
Solar System spent in the Galaxy, $t_{\rm{Gal}}=4.3\,$Gyr.  For
comparison with Fig.\,\ref{Fig:cluster}, we include the curve for
planetary perturbations (solid black curve) and the two objects Sedna
and 2012VP$_{113}$.

Recently the Solar System had a close encounter with the $M \simeq
0.15$\,\MSun\, binary star WISE J072003.20-084651.2 \citep[nicknamed
  ``Scholz's star'' after its discoverer][]{2014A&A...561A.113S} at a
distance of $q = 0.25^{+0.11}_{-0.07}$\,pc
\citep{2015ApJ...800L..17M}.  We estimate the perturbation by this
encounter using the impulse approximation
\citep{1975MNRAS.173..729H,1976BAICz..27...92R}, in which the duration
of the encounter is assumed to be much shorter that the period of the
perturbed orbit (opposite to the adiabatic regime in which
Eq.~\ref{Eq:penetration} is valid). The impulsive approximation is
fulfilled for the encounter with Scholz's star, which had a relative
velocity of $v_{\rm enc} = 83.2$\,km/s \citep{2015ApJ...800L..17M}.
We indicate this distance by the yellow curve in
Fig.\,\ref{Fig:galaxy}.

According to our analysis such an encounter (at $q_{\rm enc} \sim 5.2
\cdot 10^4$\,AU) should occur $\sim 5000$ times during the $\sim
4.3$\,Gyr sojourn of the Solar System through the Galactic disk.  Such
encounter is therefore quite a likely event, which occurs roughly once
every million years.  The close approach of Scholz's star occurred
only 70,000 years ago \citep{2015ApJ...800L..17M}, which seems
amazingly recent.  If we naively divide the two time scales, such an
encounter should already have happened $\sim 60,000$ times.

The effect of this particular encounter has hardly perturbed the
\cite{Oort1927} cloud down to a distance of $10^5$\,AU from the
Sun. But an encounter with an equally low-mass star three orders of
magnitude closer in would have affected the outer most planets.  With
the derived probability distribution Eq.\,\ref{Eq:Gamma} using
Eq.\,\ref{Eq:GravFocusCrosssection} for the cross section and
Eq.\,\ref{Eq:impulse} to estimate the distance to which such an
encounter affects the solar system.  Such an encounter would be very
unlikely to happen over the lifetime of the Solar System.

Considering the analysis of \cite{Jilkova2015} the encounter that
introduced the Sednitos (a family of planetesimals with orbits similar
to the 2003VB$_{12}$ Sedna) into the Solar System occurred in the
parental cluster, and since then no other stellar encounter has
perturbed Edgewordt-Kuiper
\citep{1943JBAA...53..181E,1951astr.conf..357K} belt. This picture is
consistent with the presence of a {\em Parking zone} (see
\S\,\ref{Sect:zones}) in the Solar System between about 100\,AU and
1000\,AU; encounters that affect the Solar System to a distance within
1000\,AU are extremely rare.

The Parking zone in the Solar System is populated by the Sednitos
\citep{Jilkova2015}; planetesimals that share a common argument of
pericenter, inclination and perihelion distance. We argue that the
Parking zone in the Solar System extends from about 100\,AU to the
about $\sim 1000$\,AU, near the outer boundary of the inner Oort
cloud.  The distribution of orbital parameters of planetesimals
discovered in this regime bear the information of the last strong
encounter the Sun experienced from the time when it was part of its
parental cluster.

\section{Perturbing planetary systems in general}\label{Sect:General}

\begin{figure}
\begin{center}
\includegraphics[width=0.5\textwidth]{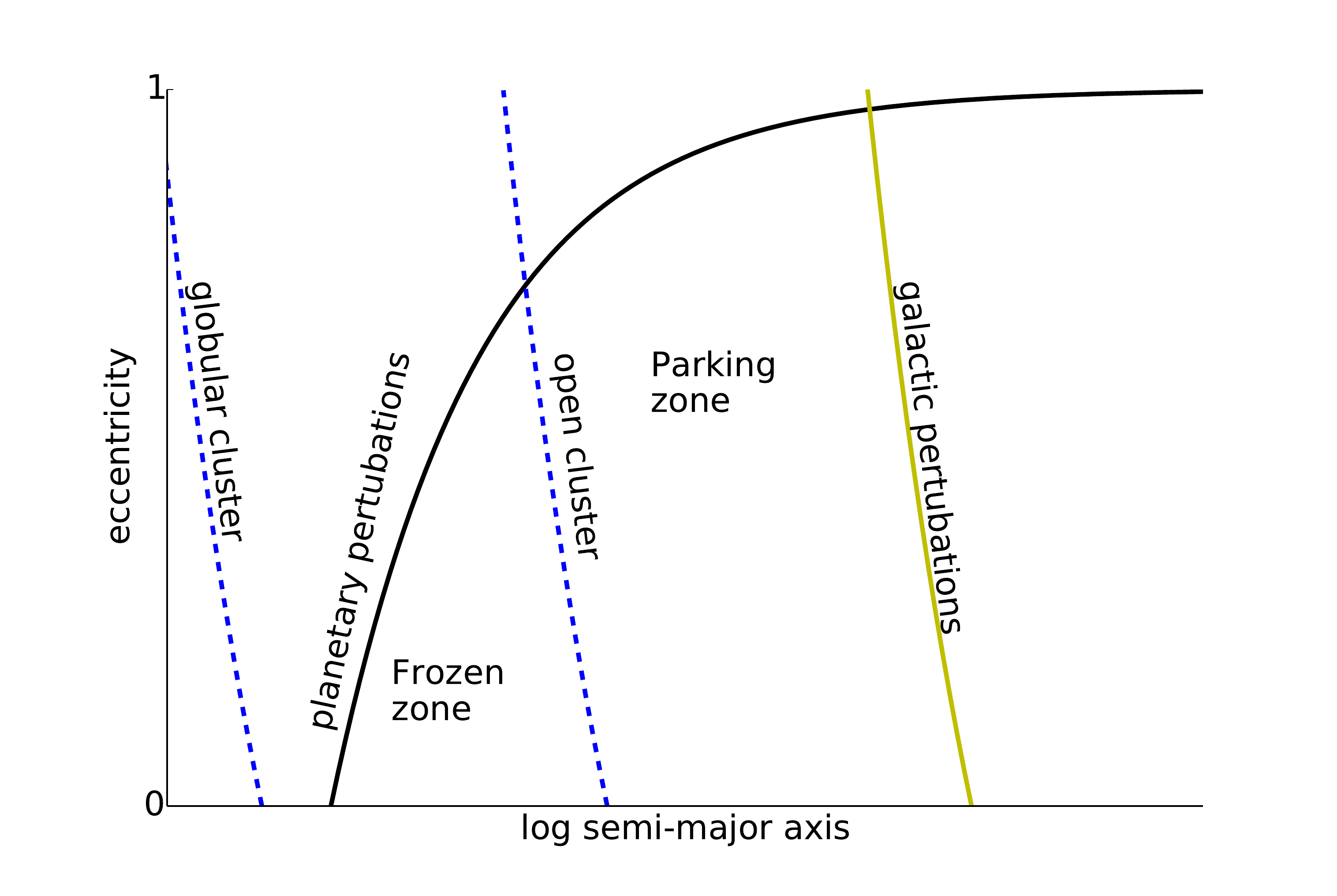}
\end{center}
\caption{The fragility of a planetary system under the influence of
  interaction with stars in its parental cluster and a galaxy.  The
  horizontal axis has no scale, because the range to which the effect
  penetrates depends on the parameters of the planetary system and the
  environment.  The black, yellow and blue curves give the ranges to
  which various perturbations affect the orbits of the minor
  bodies. Specific example of the Solar System are presented in
  Figs.\,\ref{Fig:cluster} and \ref{Fig:galaxy}.  We identify the {\em
    Frozen zone} and the {\em Parking zone}.
\label{Fig:fragility}}
\end{figure}

In Fig.\,\ref{Fig:fragility} we present a generalized view of the
fragility of planetary systems.  The solid black curve gives the range
to which possible massive planets perturb the inner parts of the
planetary system.  Violent planet scattering can cause massive planets
to migrate further outwards, where they can perturb the local
planetesimals \citep{2008ApJ...686..580C}. In that case, the solid
curve will shift to the right.

The parental star cluster perturbs its planetary systems, but the
further evolution of the planetary system determines to what range
such a perturbation is preserved over time.  For a globular cluster
the stellar density is generally higher than for an open cluster, and
the probability of spending a prolonged period in a globular cluster
is also larger. As a result the range to which the planetary system is
perturbed when born in a globular cluster is much closer to the star
than for an open cluster. In fact, from the schematic picture
(Fig.\,\ref{Fig:fragility}) the influence of random encounters while
the member of a globular clusters penetrates all the way to the inner
planets. As a consequence, planetary systems in globular clusters, or
other massive dense star clusters, are likely to be perturbed by
internal as well as external effects. This may explain, in part, the
lack of observed planets in globular clusters
\citep{2007ASPC..366..289W}.

If born in a low density cluster with a relatively short lifetime, the
range to which random encounters penetrate into the planetary system
hardly reaches the influence range of the giant planets. Once the
planetary system escapes the star cluster, the perturbations from the
Galaxy start to affect the orbits of the objects.  This influence
prolongs for the remainder of the main-sequence lifetime of the parent
star, after which stellar evolution starts to play a major role in the
redistribution of the orbits.

\subsection{The Frozen and Parking zones}\label{Sect:zones}

\begin{verse}
{\em 
We define the Parking zone as a range in semi-major axis and
eccentricity in which the orbits of objects have only been affected by
encounters in the parental star cluster, and not by the local planets
or by the Galaxy.
}
\end{verse}
Objects that orbit in the Parking zone
have therefore not been affected by the planets and remain unaffected
by close Galactic encounters. The Parking zone is likely to shrink
with time, and young stars tend to have a more extended Parking zone
than older stars, due to the less prolonged exposure to Galactic
encounters.

Objects with orbital parameters in the Parking zone preserve
information about the last event that affected their orbits in the
planetary system.  Once in the Parking zone, orbital parameters are
unlikely to be affected either by the planets, because they only
affect orbits closer to the star, or by random Galactic encounters,
because they tend to affect the outer most regions of the planetary
system.  Objects found in the Parking zone can therefore be used as
tracers to reconstruct the event that introduced them in their current
orbits.

To the left of the Parking zone, and to the right of the range to
which planets perturb the orbits we recognize the {\em Frozen zone}.
\begin{verse}
{\em We define the Frozen zone as a range in semi-major axis and
  eccentricity in which the orbits of objects have not been affected
  by the local planets and not by any encounters, in the parental star
  cluster or the Galaxy.} 
\end{verse}
In this zone minor bodies remain unaffected
by either internal influences or external perturbations. Planetesimals
found in this regime will preserve information about the formation of
the planetary system.

In the Solar System, the Frozen zone is probably very small or
completely absent (see also Fig.\,\ref{Fig:cluster}). But other
planetary system may have a populated Frozen zone, which can be used
to study their origin.

\section*{Acknowledgments}

We thank Anthony Brown for discussions.  We thank the anonymous
referee for spotting a mistake in our initial analysis.  This work was
supported by the Netherlands Research Council NWO (grants
\#643.200.503, \#639.073.803 and \#614.061.608) by the Netherlands
Research School for Astronomy (NOVA). The numerical computations were
carried out on the Little Green Machine at Leiden University.


\begin{thebibliography}{}

\bibitem[\protect\astroncite{{Adams}}{2010}]{2010ARA&A..48...47A}
{Adams}, F.~C. 2010, \araa, 48, 47

\bibitem[\protect\astroncite{{Binney} \&
  {Tremaine}}{1987}]{1987gady.book.....B}
{Binney}, J., {Tremaine}, S. 1987,
\newblock Galactic dynamics,
\newblock Princeton, NJ, Princeton University Press, 1987, 747 p.

\bibitem[\protect\astroncite{{Bouwman} et~al.}{2008}]{2008ApJ...683..479B}
{Bouwman}, J., {Henning}, T., {Hillenbrand}, L.~A., {Meyer}, M.~R., {Pascucci},
  I., {Carpenter}, J., {Hines}, D., {Kim}, J.~S., {Silverstone}, M.~D.,
  {Hollenbach}, D., {Wolf}, S. 2008, \apj, 683, 479

\bibitem[\protect\astroncite{{Brasser} et~al.}{2012}]{2012Icar..217....1B}
{Brasser}, R., {Duncan}, M.~J., {Levison}, H.~F., {Schwamb}, M.~E., {Brown},
  M.~E. 2012, \icarus, 217, 1

\bibitem[\protect\astroncite{{Brown} et~al.}{2004}]{2004ApJ...617..645B}
{Brown}, M.~E., {Trujillo}, C., {Rabinowitz}, D. 2004, \apj, 617, 645

\bibitem[\protect\astroncite{{Chatterjee} et~al.}{2008}]{2008ApJ...686..580C}
{Chatterjee}, S., {Ford}, E.~B., {Matsumura}, S., {Rasio}, F.~A. 2008, \apj,
  686, 580

\bibitem[\protect\astroncite{{Davies} et~al.}{2006}]{2006MNRAS.370.2038D}
{Davies}, M.~B., {Bate}, M.~R., {Bonnell}, I.~A., {Bailey}, V.~C., {Tout},
  C.~A. 2006, \mnras, 370, 2038

\bibitem[\protect\astroncite{{Edgeworth}}{1943}]{1943JBAA...53..181E}
{Edgeworth}, K.~E. 1943, Journal of the British Astronomical Association, 53,
  181

\bibitem[\protect\astroncite{{Fern{\'a}ndez}}{1997}]{1997Icar..129..106F}
{Fern{\'a}ndez}, J.~A. 1997, \icarus, 129, 106

\bibitem[\protect\astroncite{{Galilei}}{1632}]{1632Dialogo...G}
{Galilei}, G. 1632,
\newblock {Dello Studio di Pisa}, p.~189

\bibitem[\protect\astroncite{{Heggie}}{1975}]{1975MNRAS.173..729H}
{Heggie}, D.~C. 1975, \mnras, 173, 729

\bibitem[\protect\astroncite{{Heggie} \& {Rasio}}{1996}]{1996MNRAS.282.1064H}
{Heggie}, D.~C., {Rasio}, F.~A. 1996, \mnras, 282, 1064

\bibitem[\protect\astroncite{{Holmberg} et~al.}{2007}]{2007A&A...475..519H}
{Holmberg}, J., {Nordstr{\"o}m}, B., {Andersen}, J. 2007, \aap, 475, 519

\bibitem[\protect\astroncite{{Howard}}{2013}]{2013Sci...340..572H}
{Howard}, A.~W. 2013, Science, 340, 572

\bibitem[\protect\astroncite{{Ida} et~al.}{2013}]{2013ApJ...775...42I}
{Ida}, S., {Lin}, D.~N.~C., {Nagasawa}, M. 2013, \apj, 775, 42

\bibitem[\protect\astroncite{{Jilkova} et~al.}{2015}]{Jilkova2015}
{Jilkova}, L., {Portegies Zwart}, S., {Pijloo}, T., {Hammer}, M. 2015,
\newblock in MNRAS, Vol. Submitted

\bibitem[\protect\astroncite{{Kaib} \& {Quinn}}{2008}]{2008Icar..197..221K}
{Kaib}, N.~A., {Quinn}, T. 2008, \icarus, 197, 221

\bibitem[\protect\astroncite{{Kobayashi} \& {Ida}}{2001}]{2001Icar..153..416K}
{Kobayashi}, H., {Ida}, S. 2001, Icarus, 153, 416

\bibitem[\protect\astroncite{{Kokubo} \& {Ida}}{2002}]{2002ApJ...581..666K}
{Kokubo}, E., {Ida}, S. 2002, \apj, 581, 666

\bibitem[\protect\astroncite{{Kuiper}}{1951}]{1951astr.conf..357K}
{Kuiper}, G.~P. 1951,
\newblock in J.~A. {Hynek} (ed.), 50th Anniversary of the Yerkes Observatory
  and Half a Century of Progress in Astrophysics,  357

\bibitem[\protect\astroncite{{Lada} \& {Lada}}{2003}]{2003ARA&A..41...57L}
{Lada}, C.~J., {Lada}, E.~A. 2003, \araa, 41, 57

\bibitem[\protect\astroncite{{Lamers} et~al.}{2005}]{2005A&A...429..173L}
{Lamers}, H.~J.~G.~L.~M., {Gieles}, M., {Portegies Zwart}, S.~F. 2005, \aap,
  429, 173

\bibitem[\protect\astroncite{{Levison} et~al.}{2008}]{2008Icar..196..258L}
{Levison}, H.~F., {Morbidelli}, A., {Van Laerhoven}, C., {Gomes}, R.,
  {Tsiganis}, K. 2008, \icarus, 196, 258

\bibitem[\protect\astroncite{{Li} \& {Adams}}{2015}]{2015MNRAS.448..344L}
{Li}, G., {Adams}, F.~C. 2015, \mnras, 448, 344

\bibitem[\protect\astroncite{{Mamajek} et~al.}{2015}]{2015ApJ...800L..17M}
{Mamajek}, E.~E., {Barenfeld}, S.~A., {Ivanov}, V.~D., {Kniazev}, A.~Y.,
  {V{\"a}is{\"a}nen}, P., {Beletsky}, Y., {Boffin}, H.~M.~J. 2015, \apjl, 800,
  L17

\bibitem[\protect\astroncite{Oort}{1927}]{Oort1927}
Oort, J. 1927, ban, 3, 275

\bibitem[\protect\astroncite{{Portegies Zwart}}{2009}]{2009ApJ...696L..13P}
{Portegies Zwart}, S.~F. 2009, \apjl, 696, L13

\bibitem[\protect\astroncite{{Portegies Zwart}
  et~al.}{2010}]{2010ARA&A..48..431P}
{Portegies Zwart}, S.~F., {McMillan}, S.~L.~W., {Gieles}, M. 2010, \araa, 48,
  431

\bibitem[\protect\astroncite{{Portegies Zwart}
  et~al.}{2001}]{2001MNRAS.321..199P}
{Portegies Zwart}, S.~F., {McMillan}, S.~L.~W., {Hut}, P., {Makino}, J. 2001,
  \mnras, 321, 199

\bibitem[\protect\astroncite{{Rickman}}{1976}]{1976BAICz..27...92R}
{Rickman}, H. 1976, Bulletin of the Astronomical Institutes of Czechoslovakia,
  27, 92

\bibitem[\protect\astroncite{{Scholz}}{2014}]{2014A&A...561A.113S}
{Scholz}, R.-D. 2014, \aap, 561, A113

\bibitem[\protect\astroncite{{Schwamb}}{2014}]{2014Natur.507..435S}
{Schwamb}, M.~E. 2014, \nat, 507, 435

\bibitem[\protect\astroncite{{Steinhausen} \&
  {Pfalzner}}{2014}]{2014A&A...565A..32S}
{Steinhausen}, M., {Pfalzner}, S. 2014, \aap, 565, A32

\bibitem[\protect\astroncite{{Trujillo} \&
  {Sheppard}}{2014}]{2014Natur.507..471T}
{Trujillo}, C.~A., {Sheppard}, S.~S. 2014, \nat, 507, 471

\bibitem[\protect\astroncite{{Veras} et~al.}{2011}]{2011MNRAS.417.2104V}
{Veras}, D., {Wyatt}, M.~C., {Mustill}, A.~J., {Bonsor}, A., {Eldridge}, J.~J.
  2011, \mnras, 417, 2104

\bibitem[\protect\astroncite{{Weldrake} et~al.}{2007}]{2007ASPC..366..289W}
{Weldrake}, D.~T.~F., {Sackett}, P.~D., {Bridges}, T.~J. 2007,
\newblock in C. {Afonso}, D. {Weldrake}, T. {Henning} (eds.), Transiting
  Extrapolar Planets Workshop, Vol. 366 of {\em Astronomical Society of the
  Pacific Conference Series\/},  289

\bibitem[\protect\astroncite{{Zahn}}{1977}]{1977A&A....57..383Z}
{Zahn}, J.-P. 1977, \aap, 57, 383

\end{thebibliography}
\end{document}